\documentclass[twocolumn,english,pra,showpacs,twocolumn]{revtex4}
\usepackage{amsmath}
\usepackage{graphicx}
\usepackage{amssymb}
\usepackage{txfonts}

\begin{document}

\title{Quantum Maxwell's Demon in Thermodynamic Cycles}

\author{H. Dong, D.Z. Xu and C.P. Sun}

\email{suncp@itp.ac.cn}

\homepage{http://power.itp.ac.cn/ suncp/index.html}

\affiliation{Institute of Theoretical Physics, Chinese Academy of Sciences, Beijing,
100190, China}
\begin{abstract}
We study the physical mechanism of Maxwell's Demon (MD) helping to
do extra work in thermodynamic cycles, by describing measurement of
position, insertion of wall and information erasing of MD in a quantum
mechanical fashion. The heat engine is exemplified with one molecule
confined in an infinitely deep square potential inserted with a movable
solid wall, while the MD is modeled as a two-level system (TLS) for
measuring and controlling the motion of the molecule. It is discovered
that the the MD with quantum coherence or on a lower temperature than
that of the heat bath of the particle would enhance the ability of
the whole work substance formed by the system plus the MD to do work
outside. This observation reveals that the role of the MD essentially
is to drive the whole work substance being off equilibrium, or equivalently
working with an effective temperature difference. The elaborate studies
with this model explicitly reveal the effect of finite size off the
classical limit or thermodynamic limit, which contradicts the common
sense on Szilard heat engine (SHE). The quantum SHE's efficiency is
evaluated in detail to prove the validity of second law of thermodynamics. 
\end{abstract}


\maketitle

\section{Introduction}

Maxwell's demon (MD) has been a notorious being since its existence
could violate the second law of thermodynamics (SLoT) \cite{MD_book,Nori2009}:
the MD distinguishes the velocities of the gas molecules, and then
controls the motions of molecules to create a difference of temperatures
between the two domains. In 1929, Leo Szilard proposed the {}``one
molecular heat engine''(we call Szilard heat engine(SHE)) \cite{Szilard1929}
as an alternative version of heat engine assisted by MD. The MD firstly
measures which domain, the single molecule stays in and then gives
a command to the system for extracting work according to the measurement
results. In a thermodynamic cycle, the molecule seems to extract heat
from a single heat bath at temperature $T$, and thus do work $k_{\mathrm{B}}T\ln2$
without evoking other changes. This consequence obviously violates
the SLoT.

The first revival of the studies of MD is due to the recognition about
the trade-off between information and entropy in the MD-controlled
thermodynamic cycles. The milestone discovery is the {}``Landauer
principle'' \cite{Landauer1961}, which reveals that erasing one
bit information from the memory in computing process would inevitably
accompany an increasing entropy of the environment. In the SHE, the
erasing needs work $k_{\mathrm{B}}T\ln2$ done by the external agent.
It gives a conceptual solution for the MD paradox \cite{Bennett1982}
by considering the MD as a part of the whole work substance, thus
the erasing information stored in the demon's memory is necessary
to restart a new thermodynamic cycle. This observation about erasing
the information of the MD finally saves the SLoT.

The recent revival of the studies of MD is due to the development
of quantum information science. The corresponding quantum thermodynamics
concerns the quantum heat engines (QHEs) \cite{Kieu2004,htquan} utilizing
quantum coherent system serving as the work substance. A quantum work
substance is a quantum system off the thermodynamic limit, which perseveres
its quantum coherence \cite{Scully2003,htquan2006} to some extent,
and obviously has tremendous influence on the properties of QHEs.
Especially, when quantum MD is included in the thermodynamical cycle
\cite{Zurek1984,lloyd1997,Quan2006}, the whole work substance formed
by the work substance plus MD would be off the thermodynamic limit
and possesses some quantum coherence. There are many attempts to generalize
the SHE by quantum mechanically approaching the measurement process
\cite{Zurek1984}, the motion control \cite{Quan2006}, the insertion
and the expansion process \cite{Ueda2010}. However, to our best knowledge,
a fully quantum approach for all actions in the SHE integrated with
a quantum MD intrinsically is still lack. The quantum-classical hybrid
description of the SHE may result in some notorious observations about
MD assisted thermodynamic process, which seems to challenge the common
senses in physics. Therefore, we need a fully quantum theory for the
MD-assisted thermodynamics.

In this paper, we propose a quantum SHE assisted by MD with a finite
temperature different from that of the system. In this model, we give
a consistent quantum approach to the measurement process without using
the von Neumann projection \cite{Pzhang}. Then we calculate the works
done by the insertion of the movable wall in the framework of quantum
mechanics. The controlled gas expansion is treated with the quantum
conditional dynamics. Furthermore, we also consider the process of
removing wall to complete a thermodynamic cycles. With these necessary
subtle considerations, the quantum approach for the MD-assisted thermodynamic
cycle will go much beyond the conventional theories about the SHE.
We show that the system off the thermodynamic limit exhibits uncommon
observable quantum effects due to the finite size of system , which
results in the discrete energy levels that could be washed out by
the heat fluctuation. Quantum coherence can assist the MD to extract
more work by reducing effective temperature, while thermal excitation
of the MD at a finite temperature would reduce its abilities for quantum
measurement and conditional control of the expansion. It means that,
only cooled to the absolute zero temperature, could the MD help the
molecule to do maximum work outside.

Our paper is organized as follows: In Sec. II we firstly give a brief
review of classical version of SHE, and then present our model in
quantum version with MD included intrinsically. The role of quantum
coherence of MD is emphasized with the definition of the effective
temperature for arbitrary two level system. In the Sec. III of main
part, we consider the quantum SHE with a quantum MD at finite temperature,
doing measurement for the position of the particle confined in a one
dimensional infinite deep well. The whole cycle consists four steps:
insertion, measurement, expansion and removing. Detailed descriptions
are performed subsequently in the whole cycle of SHE. We calculate
the work done and heat exchange in every sub-step. In Sec. IV, we
discuss quantum SHE's operation efficiency in comparison with the
Carnot heat engine. We restore the well-known results in the classical
case by tuning the parameters in the quantum version, such as the
width of the potential well. Conclusions and remarks are given in
Sec. VI.

\section{Quantum Maxwell's Demon in Szilard Heat Engine\label{sec:II}}

\begin{figure}
\includegraphics[bb=4bp 385bp 411bp 776bp,width=7cm]{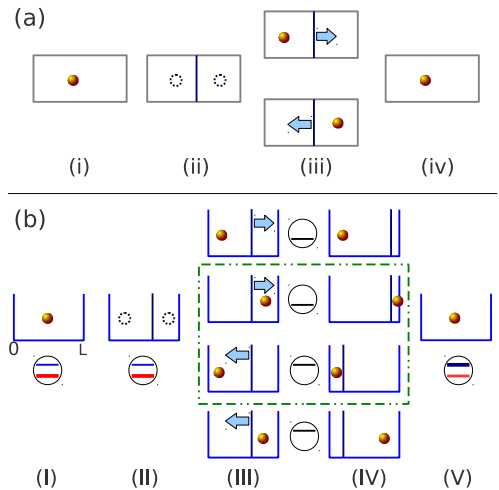} \caption{Classical and quantum Szilard's single molecular heat engine. \textbf{(a)}
Classical version: (i-ii) A piston is inserted in the center of a
chamber. (ii-iii) The demon finds which domain, the single molecule
stays in. (iii-iv) The demon controls the system to do work according
to its memory; \textbf{(b)} Quantum version: The demon is modeled
as a two level system with two energy levels $\left|g\right\rangle $
and $\left|e\right\rangle $ and energy spacing $\Delta$. The chamber
is quantum mechanically described as an infinite potential with width
$L$. (I-II) An impenetrable wall is inserted at arbitrary position
in the potential. (II-III) The demon measures the state of the system
and then record the results in its memory by flipping its own state
or no action taken. The measurement may result in the wrong results
illustrated in the green dashed rectangle. (III-IV) The demon controls
the expansion for the single molecule according to the measurement.
(IV-V) The wall is removed from the potential. }

\label{fig:cycle} 
\end{figure}

In this section, we firstly revisit Szilard's single molecular heat
engine (SHE) in brief. As illustrated in Fig. \ref{fig:cycle}(a),
the whole thermodynamic cycle consists three steps: insertion(i-ii),
measurement(ii-iii) and controlled expansion(iii-iv) by the MD. The
demon inserts a piston isothermally in the center of the chamber.
Then, it finds which domain, the single molecule stays in and changes
its own state to register the information of the system. Without losing
generality, we assume the demon initially is in the state $0$. Finding
the molecule is on the right, namely $L/2<x<L$, the demon changes
its own memory to state $1$, while it does not change if the molecule
is on the left $\left(0<x<L/2\right)$. According to the information
acquired in the measurement process, the demon controls the expansion
of the domain with the single molecule: allowing the isothermal expansion
with the piston moving from $L/2$ to $L$ if its memory registers
$0$, and moving from $L/2$ to $0$, if the register is on the state
$1$. In each thermodynamic cycle, the system does work $W=k_{\mathrm{B}}T\ln2$
to the outside agent in the isothermal expansion. In an overall looking,
the system extracts heat from a single heat bath to do work, thus
it would violate the SLoT if the MD were not treated as a part of
the work substance in the SHE. However, after the cycle, MD stores
one bit information as its final state and is in the mixture of $0$
and $1$ states with equal probability. Thus it does not return to
its initial state. Landauer's principle states that to erase such
a bit of information at temperature $T$ requires the dissipation
of energy at least $k_{\mathrm{B}}T\ln2$. The work extracted by the
system just compensates the energy for erasing the information. Therefore,
the SLoT is saved. In this sense, the classical version of MD paradox
is only a misunderstanding, due to the ignorance of some roles of
the MD \cite{Bennett1982}.

In the most of previous investigations about the MD paradox, it is
usually assumed the system and the MD possess the same heat bath.
Thus the whole work substance formed by the system plus the MD is
in equilibrium, and no quantum coherence exists. If the demon is in
contact with a lower temperature heat bath while the system's environment
possesses higher temperature $T$, the work needed in the erasing
process is smaller than $k_{B}T\ln2$ \cite{Quan2006}. Under this
circumstance, we actually construct a quantum heat engine with non-equilibrium
or an equilibrium working substance work between two different heat
baths. Furthermore, when the MD is initially prepared with quantum
coherence, the quantum nature of the whole work substance results
in many exotic functions for QHE.

To tackle this problem, we study here a quantum version of Szilard's
model with an MD accompanying it. In this model, the chamber is modeled
as an infinite square potential well with the width $L$, as illustrated
in Fig. \ref{fig:cycle}(b). And the demon is realized by a single
two-level atom with energy levels $\left\vert g\right\rangle $, $\left\vert e\right\rangle $
and level spacing $\Delta$. Initially, the system is in thermal state
with inverse temperature $\beta$. And the demon has been in contact
with the low temperature bath at the inverse temperature $\beta_{D}$.
Namely, the demon is initial prepared in the equilibrium state

\begin{equation}
\rho_{D}=p_{g}\left\vert g\right\rangle \left\langle g\right\vert +p_{e}\left\vert e\right\rangle \left\langle e\right\vert ,\end{equation}
 with the probability $p_{e}=1-p_{g}$ in the excited state and one
in ground state \[
p_{g}=1/\left[1+\exp\left(-\beta_{D}\Delta\right)\right].\]

Actually, the inverse temperature $\beta_{D}$ could represent an
effective inverse temperature of the MD with quantum coherence. For
an environment being a mesoscopic system, the number of its degrees
of freedom is not so large. Under this circumstance, the strong coupling
to the MD leaves finite off-diagonal elements in the reduced density
matrix\cite{Dong2007}. This remnant of coherence can be utilized
to improve the apparent efficiency of the heat engine \cite{Scully2003,htquan2006}.
For the demon with coherence, the density matrix usually reads as
\begin{equation}
\rho_{D}=\left[\begin{array}{cc}
p_{g} & F\\
F^{\ast} & p_{e}\end{array}\right],\label{eq:cohe}\end{equation}
 where the off-diagonal element $F$ measures the quantum coherence.
The eigen-values of the above reduced density matrix represent two
effective population probabilities as\begin{eqnarray}
p_{+}\left(F\right) & \simeq & p_{e}-\coth\left(\frac{\Delta}{2}\beta_{D}\right)\left\vert F\right\vert ^{2},\notag\\
p_{-}\left(F\right) & \simeq & p_{g}+\coth\left(\frac{\Delta}{2}\beta_{D}\right)\left\vert F\right\vert ^{2}.\end{eqnarray}
 We can define an effective inverse temperature $\beta_{\mathrm{eff}}=\ln p_{+}\left(F\right)/p_{-}\left(F\right)$
for the two-level MD, namely, \begin{equation}
\beta_{\mathrm{eff}}=\beta_{D}+\frac{4\left\vert F\right\vert ^{2}}{\Delta}\cosh^{2}\left(\frac{\Delta}{2}\beta_{D}\right)\coth\left(\frac{\Delta}{2}\beta_{D}\right).\end{equation}
 The effective temperature $T_{\mathrm{eff}}=1/\beta_{\mathrm{eff}}$
here is lower than the bath temperature $T_{D}$. As shown as follows,
it is the lowing of the effective temperature of the MD that results
in an increasing of the heat engine efficiency.

As for the modeling of the chamber as an infinite square potential
well, the eigenfunctions of the confined single molecule are \begin{equation}
\left\langle x\right.\left\vert \psi_{n}\left(L\right)\right\rangle =\sqrt{\frac{2}{L}}\sin\left[n\pi x/L\right],\end{equation}
 with the corresponding eigen-energies $E_{n}\left(L\right)=\left(\hbar n\pi\right)^{2}/\left(2mL^{2}\right)$,
where the quantum number $n$ ranges from $1$ to $\infty$.

On this bases, the initial state of the total system is expressed
as a product state \begin{equation}
\rho_{0}=\frac{1}{Z\left(L\right)}\sum_{n}e^{-\beta E_{n}\left(L\right)}\left\vert \psi_{n}\left(L\right)\right\rangle \left\langle \psi_{n}\left(L\right)\right\vert \otimes\rho_{0}^{D},\end{equation}
 where \begin{equation}
Z\left(L\right)=\sum_{n}\exp\left[-\beta E_{n}\left(L\right)\right]\end{equation}
 is the partition function of the system.

Here, we remark that the discrete spectrum of the system results from
the finite size of the width $L.$ As $L\rightarrow\infty$, the spectrum
becomes continuous as the energy level spacings is proportional to
$1/L^{2}$. Then heat excitation characterized by $k_{B}T$ can wash
out the quantum effect so that the system approaches a classical limit.
Some of finite size effect based quantum phenomenon could also disappear
as $T\rightarrow\infty.$

With the above modelings, the MD-assisted thermodynamic cycle for
the quantum SHE is divided as four steps illustrated in Fig. \ref{fig:cycle}(b):
(I-II) the insertion of a mobile solid wall into the potential well
at a position $x=l$ (the origin is $x=0$ ); (II-III) the measurement
done by the MD to create the quantum entanglement of its two internal
states to the spatial wave functions of the confined molecule; (III-IV)
quantum control for the mobile wall to move the according to the record
in the demon's memory; (IV-V) removing the wall so that the next thermodynamic
cycle can be restarted. Their descriptions will be discussed subsequently
in the next section and detailed calculations will be found in the
Appendix.

\bigskip{}

\section{Quantum Thermodynamic Cycles with Measurement}

In this section we analyze in details the thermodynamic cycle of the
molecule confined in an infinite square potential well. The molecule's
position is monitored and then controlled by the MD. The MD may have
quantum coherence as in Eq.\ref{eq:cohe}, or equivalently, possesses
a lower temperature $T_{\mathrm{D}}=1/\beta_{\mathrm{D}}$ than $T=1/\beta$
of the confined molecule's heat bath. In each step, we will evaluate
the work done by outside agent and heat exchange in detail. In order
to concentrate on the physical properties, we put the calculations
in the Appendix.

\begin{figure}
\includegraphics[width=5cm]{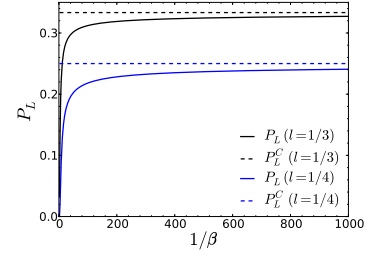} \caption{(Color Online) Probability $P_{L}$ and the corresponding classical
one $P_{L}^{C}$ vs temperature $1/\beta$ for different piston position
$l=1/3$ and $l=1/4$. Without losing generality, we set the parameters
as $L=1$, $m=\pi^{2}/2$ and $\hbar=1$.}

\label{fig:pl} 
\end{figure}

\subsection*{Step1: Quantum Insertion (I-II)}

In the first process, the system is in contact with the heat bath
$\beta$, then a piston is inserted isothermally into the potential
at position $l$. The potential is then divided into two domains,
denoted simply as $L$ and $R$, with the length $l$ and $L-l$ respectively.
The eigenstates change into the following two sets as

\begin{eqnarray}
\left\langle x\right.\left|\psi_{n}^{R}(L-l)\right\rangle  & = & \begin{cases}
\sqrt{\frac{2}{L-l}}\sin\left[\frac{n\pi\left(x-l\right)}{L-l}\right] & l\leq x\leq L\\
0 & 0\leq x\leq l\end{cases},\end{eqnarray}
 and

\[
\left\langle x\right.\left\vert \psi_{n}^{L}(l)\right\rangle =\begin{cases}
0 & l\leq x\leq L\\
\sqrt{\frac{2}{l}}\sin(n\pi x/l) & 0\leq x\leq l\end{cases},\]
 with the corresponding eigen-values $E_{n}\left(L-l\right)$ and
$E_{n}\left(l\right)$. In the following discussions we use the free
Hamiltonian $H_{T}=H+H_{D}$ for \begin{eqnarray*}
H & = & \sum_{n}[E_{n}\left(l\right)\left\vert \psi_{n}\left(l\right)\right\rangle \left\langle \psi_{n}\left(l\right)\right\vert \\
 &  & +E_{n}(L-l)\left\vert \psi_{n}\left(L-l\right)\right\rangle \left\langle \psi_{n}\left(L-l\right)\right\vert ]\end{eqnarray*}
 for $0\leq l\leq L$ and $H_{D}=\Delta\left\vert e\right\rangle \left\langle e\right\vert .$
Here, we take its ground state energy as the zero point of energy
of atom.

At the end of the insertion process, the system is still in the thermal
state with the temperature $\beta$ and the MD is on its own state
without any changes. With respect to the above splitted bases, the
state of the whole system is rewritten in terms of the new bases as

\begin{equation}
\rho_{\mathrm{ins}}=[P_{L}\left(l\right)\rho^{L}\left(l\right)+P_{R}\left(l\right)\rho^{R}\left(L-l\right)]\otimes\rho_{0}^{D},\end{equation}
 where \begin{equation}
\rho^{L}\left(l\right)=\sum_{n}\frac{e^{-\beta E_{n}\left(l\right)}}{Z\left(l\right)}\left\vert \psi_{n}^{L}(l)\right\rangle \left\langle \psi_{n}^{L}(l)\right\vert ,\end{equation}
 and \begin{equation}
\rho^{R}\left(L-l\right)=\sum_{n}\frac{e^{-\beta E_{n}\left(L-l\right)}}{Z\left(L-l\right)}\left\vert \psi_{n}^{R}(L-l)\right\rangle \left\langle \psi_{n}^{R}(L-l)\right\vert ,\end{equation}
refer to the system localized in the left and right domain respectively.
With respect to the their sum $\mathcal{Z}\left(l\right)=Z\left(l\right)+Z\left(L-l\right),$
the temperature dependent ratios \[
P_{L}\left(l\right)=Z\left(l\right)/\mathcal{Z}\left(l\right)\]
 and \[
P_{R}\left(l\right)=Z\left(L-l\right)/\mathcal{Z}\left(l\right).\]
 are the probabilities to find the single molecule on the left and
the right side respectively. For simplicity, we denote $P_{L}\left(l\right)$
and $P_{R}\left(l\right)$ by $P_{L}$ and $P_{R}$ respectively in
the following discussions. We emphasize that the probabilities are
different from the classical probabilities, $P_{L}^{c}=l/L$ and $P_{L}^{c}=\left(L-l\right)/L$,
finding single molecule on the left and right side that is proportional
to the volume. We numerically illustrate this discrepancy between
this classical result and ours in Fig. \ref{fig:pl} for different
insertion position $l=1/3$ and $l=1/4$. It is clearly in Fig. \ref{fig:pl}
that the probabilities $P_{L}$ approaches to the corresponding classical
ones $P_{L}^{c}$, as the temperature increases to the high temperature
limit. However, a large discrepancy is observed at low temperature.
This deviate from the classical one is due to the discreteness of
the energy levels of the potential well with finite width, which disappears
as level spacing becomes small with $L\rightarrow\infty$. In this
case, the heat excitation will erase all the quantum feature of the
system and the classical limit is approached.

\begin{figure}
\includegraphics[width=8cm]{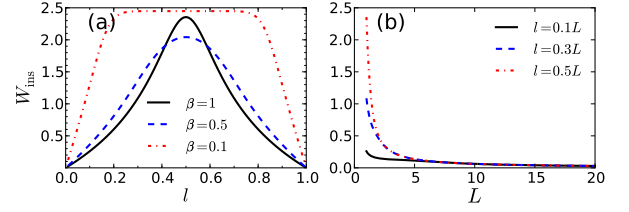} \caption{(Color Online) Work done by the outside agent. (a)$W_{\mathrm{ins}}$
vs $l$ for different system inverse temperature $\beta=1$, $0.5$
and $0.1$. Here, we choose the same parameter as that in Fig. \protect\ref{fig:pl}.
(b) $W_{\mathrm{ins}}$ vs $L$ for different insertion position $l=0.1L$,
$0.3L$ and $0.5L$.}

\label{fig:wins} 
\end{figure}

In this step, work should be done to the system. In the isothermal
process, the work done by the outside agent can be expressed as $W_{\mathrm{ins}}=\Delta U_{\mathrm{ins}}-T\Delta S_{\mathrm{ins}}$,
with the internal energy change \[
\Delta U_{\mathrm{ins}}=\mathrm{Tr}\left[\left(\rho_{\mathrm{ins}}-\rho_{0}\right)H_{T}\right]\]
 and the total entropy change \[
\Delta S_{\mathrm{ins}}=\mathrm{Tr}\left(-\rho_{\mathrm{ins}}\ln\rho_{\mathrm{ins}}+\rho_{0}\ln\rho_{0}\right).\]
 During this isothermal process, the work done by outside just compensates
the change of the free energy as \begin{equation}
W_{\mathrm{ins}}=T\left[\ln Z\left(L\right)-\ln\mathcal{Z}\left(l\right)\right].\end{equation}
 The same result has been obtained in Ref. \cite{Ueda2010}. By taken
inverse temperature $\beta=1$ and $L=1$, we illustrate the work
needed for the insertion of the piston into the potential in Fig.
\ref{fig:wins}. It is shown that to insert the piston at the center
of the potential needs the maximum work to be done. Another reasonable
fact is that no work is needed to insert the piston at position $l=0$
and $l=L$. Classically, it is well known that no work should be paid
for inserting the piston at any position, while for a fixed $L$,
we notice that $W_{\mathrm{ins}}\rightarrow-\infty$ as $T\rightarrow\infty$.
The discrete property of the system due to the finite width of the
potential well results in the typical quantum effect, even at a high
temperature, namely, $\lim_{T\rightarrow\infty}W_{\mathrm{ins}}\neq0$
and $\lim_{T\rightarrow\infty}Q_{\mathrm{ins}}\neq0$. This finite
size induced quantum effect is typical for mesoscopic system. To restore
the classical results, we simply take the limit $L\rightarrow\infty$
to make the spectrum continuous, rather than $T\rightarrow\infty$.
Under this limit $L\rightarrow\infty$, we have $\mathcal{Z}\left(l\right)/Z\left(L\right)\rightarrow1$,
which just recovers the classical result that \begin{equation}
\lim_{L\rightarrow\infty}W_{\mathrm{ins}}=0,\label{eq:asyw}\end{equation}
 as illustrated in Fig. \ref{fig:wins}(b) for different insertion
positions $l=0.1L$, $0.3L$ and $0.5L$.

After the insertion of piston, the entropy of the system changes.
The system exchanges heat with the heat bath during this isothermal
reversible process. The heat is obtained by $Q_{\mathrm{ins}}=-T\Delta S_{\mathrm{ins}}$
as \begin{equation}
Q_{\mathrm{ins}}=\left(T-\frac{\partial}{\partial\beta}\right)\left[\ln Z\left(L\right)-\ln\mathcal{Z}\left(l\right)\right].\end{equation}
 Similar to the asymptotic properties of the work in Eq. (\ref{eq:asyw}),
$Q_{\mathrm{ins}}$ approaches to zero when $L\rightarrow\infty$.

\subsection*{Step2: Quantum Measurement (II-III)}

In the second step, the system is isolated from the heat bath. The
MD finds which domain, the single molecule stays in and registers
the result into its own memory. In the classical way, the memory can
also be imaged as a chamber with single molecule. The classical state
of single molecule on the right and left side are denoted as the state
$0$ and $1$. And the memory is architecture always by two bistable
states with no energy difference $\Delta=0$ and no energy is needed
in the measurement process. This setup based on {}``chamber '' argument
seems to exclude the possibility for quantum coherence in a straightforward
way. Therefore, we adopt the TLS as the memory to allow the quantum
coherence to take the role, as discussed in Sec. \ref{sec:II}. In
the scheme here, the demon performs the controlled-NOT operation \cite{Quan2006}.
If the molecule is on the left side, no operation is done. And the
demon flips its state, when finding the molecule on the right. This
operation is realized by the following unitary operator,

\begin{eqnarray}
U & = & \sum_{n}\left\vert \psi_{n}^{L}\left(l\right)\right\rangle \left\langle \psi_{n}^{L}\left(l\right)\right\vert \otimes\left(\left\vert g\right\rangle \left\langle g\right\vert +\left\vert e\right\rangle \left\langle e\right\vert \right)\notag\\
 &  & +\left\vert \psi_{n}^{R}\left(L-l\right)\right\rangle \left\langle \psi_{n}^{R}\left(L-l\right)\right\vert \otimes\left(\left\vert e\right\rangle \left\langle g\right\vert +\mathrm{h.c}\right).\end{eqnarray}
 After the measurement, the MD and the system are correlated. This
correlation enables the MD to control the system to perform work to
the outside agent. The density matrix of the whole system after measurement
is \begin{eqnarray}
\rho_{\mathrm{mea}} & = & \left[P_{L}p_{g}\rho^{L}\left(l\right)+P_{R}p_{e}\rho^{R}\left(L-l\right)\right]\otimes\left\vert g\right\rangle \left\langle g\right\vert \notag\\
 &  & +\left[P_{L}p_{e}\rho^{L}\left(l\right)+P_{R}p_{g}\rho^{R}\left(L-l\right)\right]\otimes\left\vert e\right\rangle \left\langle e\right\vert .\end{eqnarray}
 If the temperature of the demon is zero, namely $T_{D}=0$, the measurement
actually results in a perfect correlation between the system and the
MD, \begin{equation}
\rho_{\mathrm{mea}}=P_{L}\rho^{L}\left(l\right)\otimes\left\vert g\right\rangle \left\langle g\right\vert +P_{R}\rho^{R}\left(L-l\right)\otimes\left\vert e\right\rangle \left\langle e\right\vert .\end{equation}
 Then the demon can distinguish exactly the domain where the single
molecule stays, e.g. state $\left\vert g\right\rangle $ representing
the molecule on left side and vice visa. At a finite temperature,
this correlation gets ambiguous. As illustrated in the dashed green
box in Fig. \ref{fig:cycle}(b), the demon actually gets the wrong
information about the domain, where the single molecule stays. For
example, the demon thinks the molecule is on the left with memory
registering $\left\vert g\right\rangle $, while the molecule is actually
on the right. The MD loses a certain amount of information about the
system and lowers its ability to extract work. \ For case $\Delta\neq0$
at finite temperature, the above imperfect correlation leads to a
condition for the MD's temperature, \ under which \ the total system
could extract positive work.

The worst case is that, when we first let the MD to become degenerate,
i.e., $\Delta=0,$ then the temperature to approach zero. In this
sense the demon is prepared in s mixing state \[
\rho_{0}^{D}\left(\Delta=0\right)=\frac{1}{2}\left(\left\vert g\right\rangle \left\langle g\right\vert +\left\vert e\right\rangle \left\langle e\right\vert \right)\]
 and the state of the whole system after the measurement reads \begin{equation}
\rho_{\mathrm{mea}}=\left[\rho^{L}\left(l\right)+\rho^{R}\left(L-l\right)\right]\otimes\rho_{0}^{D}\left(\Delta=0\right).\end{equation}
 Thus, no information is obtained by MD. There exists another limit
process that the non-degenerate MD is firstly prepared in the zero-temperature
environment, and then let $\Delta$ approach zero. Thus, the state
of the MD is broken into $\left\vert g\right\rangle \left\langle g\right\vert $
of $\rho_{0}^{D}\left(\Delta=0\right)$. In this case, we get a more
cleaver MD as mentioned above. The physical essence of the difference
between the two limit processes lies on the symmetry breaking\cite{jqliao2009}
(we will discuss this again later). With such symmetry breaking, the
degenerate MD could also make an ideal measurement. A intuitive understanding
for the zero-temperature MD helping to do work is that a more calm
MD can see the states of the molecule more clear, thus control it
more effectively.

Next we calculate the work done in the measurement process by assuming
the total system being isolated from the heat bath of the molecule.
The heat exchange here is exactly zero, namely $Q_{\mathrm{mea}}=0$,
since the operation is unitary and the total entropy is not changed
during this process. However, the total internal energy changes, which
merely results from the work done by the outside agent\[
W_{\mathrm{mea}}=P_{R}\left(p_{g}-p_{e}\right)\Delta.\]
 to register the MD's memory. The work needed is actually a monotonous
function of the demon's bath temperature $T_{D}$. If the temperature
of the demon is zero ( the MD is prepared in a pure state ), namely
$T_{D}=0$, the work reaches its maximum $W_{\mathrm{mea}}^{\mathrm{max}}=P_{R}\Delta$.
The demon can distinguish exactly the domain, where the single molecule
stays, state $\left\vert g\right\rangle $ representing molecule on
the left, and vice visa. As discussed as follows, the work done by
the outside agent here is the main factor to low down the efficiency
of the heat engine. However, the low temperature results in a more
perfect quantum correlation between the MD and the system, thus enables
the MD to extract more work. Requirement of the work done in the measurement
and the ability of controlling free expansion are two competing factors
of the QHE. Finally, we prove that a low temperature of the demon
results in the high efficiency of quantum heat engine in the following
section. It is clear that less work is needed, if the insertion position
is closer to the right boundary of the potential. And the work needed
in the measurement process approaches to zero, namely $W_{mea}\rightarrow0$,
when $l\rightarrow L$. Thus, the efficiency is promoted to reach
the corresponding Carnot efficiency when $l=L$ for this measurement.

\subsection*{Step3: Controlled Expansion (III-IV)}

In the third step, the system is brought into contact with the heat
bath with temperature $\beta$. Then the expansion is performed slowly
enough to enable the process to be reversible and isothermal. The
expansion is controlled by the demon according to its memory. Finding
its state on $\left\vert g\right\rangle $, the outside agent allows
the piston to move right, thus the single molecule performs work to
the outside. However, the agent pays some work to move piston to the
right if the MD's memory is inaccurate, e.g. the situation in the
green dashed box in Fig. \ref{fig:cycle}(b). If in state $\left\vert e\right\rangle $,
the piston is allowed to move to the left side. Under this description,
we avoid the conventional heuristic discussion with adding an object
in the classical version of SHE. Here, we choose two arbitrary final
positions of the controlled expansion as $l_{g}$ and $l_{e}$ for
the corresponding MD's state $\left\vert g\right\rangle $ and $\left\vert e\right\rangle $.
We will prove later that the total work extracted is independent on
the expansion position chosen here. After the expansion process, the
density matrix of the whole system is expressed as \begin{eqnarray}
\rho_{\mathrm{exp}} & = & \left[P_{L}p_{g}\rho^{L}\left(l_{g}\right)+P_{R}p_{e}\rho^{R}\left(L-l_{g}\right)\right]\otimes\left\vert g\right\rangle \left\langle g\right\vert \notag\\
 &  & +\left[P_{L}p_{e}\rho^{L}\left(l_{e}\right)+P_{R}p_{g}\rho^{R}\left(L-l_{e}\right)\right]\otimes\left\vert e\right\rangle \left\langle e\right\vert .\end{eqnarray}
 During the expansion, the system performs work $-W_{\mathrm{exp}}\geq0$
to the outside agent, \begin{eqnarray}
W_{\mathrm{exp}} & = & T\left[\ln\mathcal{Z}\left(l\right)+P_{L}\ln P_{L}+P_{R}\ln P_{R}\right.\notag\\
 &  & -P_{L}p_{g}\ln Z\left(l_{g}\right)-P_{R}p_{e}\ln Z\left(L-l_{g}\right)\notag\\
 &  & \left.-P_{L}p_{e}\ln Z\left(l_{e}\right)-P_{R}p_{g}\ln Z\left(L-l_{e}\right)\right].\end{eqnarray}
 For a perfect correlation ($p_{g}=1$), the piston is moved to the
side of the potential, namely $l_{g}=L$ and $l_{e}=0$, and the work
is simply \[
W_{\mathrm{exp}}=T\left(P_{L}\ln P_{L}+P_{R}\ln P_{R}\right)-W_{\mathrm{ins}},\]
 which is the maximum work one can be extracted in this process. In
the classical limit $L\rightarrow\infty$, and the work is \[
W_{\mathrm{exp}}=T\left(P_{L}\ln P_{L}+P_{R}\ln P_{R}\right).\]
 We restore the well known result $W_{\mathrm{exp}}=-k_{\mathrm{B}}T\ln2$,
when the piston is inserted in the center of the potential. If the
demon is not perfectly correlated to the position of the single molecule
($p_{g}<1$), the work extracted $-W_{\mathrm{exp}}$ would be less.
Therefore, it is clear that the ability of MD to extract work closely
depends on the accuracy of the measurement.

In this step, the heat exchange is related to the change of entropy
as \begin{align}
\!\!\!\! Q_{\mathrm{exp}}= & P_{L}\left(T-\frac{\partial}{\partial\beta}\right)\left[\ln Z\left(l\right)-p_{g}\ln Z\left(l_{g}\right)-p_{e}\ln Z\left(L-l_{e}\right)\right]\notag\\
 & \!\!\!\!+P_{R}\left(T-\frac{\partial}{\partial\beta}\right)\left[\ln Z\left(L-l\right)-p_{e}\ln Z\left(L-l_{g}\right)-p_{g}\ln Z\left(l_{e}\right)\right].\end{align}

\subsection*{Step4: Removing(IV-V)}

To complete the thermodynamic cycle, the system and the MD should
be reset to their own initial states respectively. As for the system,
the piston inserted in the first step should be removed. In the previous
studies, this process is neglected, since the measurements are always
ideal and the piston is moved to an end boundary of the chamber. Thus
no work is required to remove piston. However, in an arbitrary process,
we can show the importance of removing piston in the whole cycle.
During this process, the system keeps contact with the heat bath with
inverse temperature $\beta$ and the removing is performed isothermally.
The density matrix of the total system after removing the piston reads
\begin{eqnarray}
\!\!\!\!\!\!\!\!\rho_{\mathrm{rev}} & = & \sum_{n}\frac{e^{-\beta E_{n}\left(L\right)}}{Z(L)}\left\vert \psi_{n}\left(L\right)\right\rangle \left\langle \psi_{n}\left(L\right)\right\vert \otimes\notag\\
 &  & \!\!\!\!\!\left[\left(P_{L}p_{g}\!+\! P_{R}p_{e}\right)\left\vert g\right\rangle \!\!\left\langle g\right\vert \!+\!\left(P_{L}p_{e}\!+\! P_{R}p_{g}\right)\left\vert e\right\rangle \!\!\left\langle e\right\vert \right].\end{eqnarray}
 In this process, the work done and the heat absorbed by the outside
are \begin{eqnarray}
\!\!\!\!\!\!\! W_{\mathrm{rev}} & = & \mathrm{Tr}\left[\left(\rho_{\mathrm{rev}}-\rho_{\mathrm{exp}}\right)\left(H+H_{D}\right)\right]\notag\\
 &  & -T\mathrm{Tr}\left[-\rho_{\mathrm{rev}}\ln\rho_{\mathrm{rev}}\right]+T\mathrm{Tr}\left[-\rho_{\mathrm{exp}}\ln\rho_{\mathrm{exp}}\right],\end{eqnarray}
 and \begin{equation}
Q_{\mathrm{rev}}=-T\mathrm{Tr}\left[-\rho_{\mathrm{rev}}\ln\rho_{\mathrm{rev}}\right]+T\mathrm{Tr}\left[-\rho_{\mathrm{exp}}\ln\rho_{\mathrm{exp}}\right],\end{equation}
 respectively. We refer the Appendix for the exact expression of those
two formula. The MD now is no longer entangled with the system. And
the density matrix of the demon is factorized out as \begin{equation}
\rho_{\mathrm{rev}}^{D}=\left(P_{L}p_{g}\!+\! P_{R}p_{e}\right)\left\vert g\right\rangle \!\!\left\langle g\right\vert \!+\!\left(P_{L}p_{e}\!+\! P_{R}p_{g}\right)\left\vert e\right\rangle \!\!\left\langle e\right\vert .\end{equation}
 In the ideal case $T_{D}=0$, the demon is on the state\[
\rho_{\mathrm{rev}}^{D}=P_{L}\left\vert g\right\rangle \left\langle g\right\vert +P_{e}\left\vert e\right\rangle \left\langle e\right\vert \]
 with entropy \[
S_{\mathrm{rev}}^{D}=-P_{L}\ln P_{L}-P_{R}\ln P_{R}.\]
 According to Landauer's Principal, erasing the memory of the MD dissipates
at least $T_{D}S_{\mathrm{rev}}^{D}=0$ work into the environment.
In this sense, we can extracted $k_{B}T\ln2$ work with MD's help.
However, we does not violate the SLoT, since the whole system functionalizes
as a heat engine working between high temperature bath and zero temperature
bath. Actually, the increase of entropy in the zero temperature bath
is exactly $S_{\mathrm{rev}}^{D}$. Therefore, the energy dissipated
actually depends on the temperature of environment, where the information
is erased. In the previous studies, people always set the same temperature
for the system and MD. Thus the exactly mechanism of MD was not clear
to certain extent, especially for SHE. Let's consider another special
case $\Delta=0$, which directly results in $p_{e}=p_{g}=1/2$. MD
is prepared on its maximum entropy state $\rho_{0}^{D}\left(\Delta=0\right)$.
At the end of the cycle, MD actually is on the same state, namely
$\rho_{\mathrm{rev}}^{D}=\rho_{0}^{D}\left(\Delta=0\right)$. Thus,
no work is paid to erase the memory.

After this procedure, the MD is decoupled from the system and brought
into contact with its own thermal bath with inverse temperature $\beta_{D}$.
Since \begin{equation}
P_{L}p_{e}+P_{R}p_{g}\geq p_{e},\end{equation}
 the MD releases energy into its heat bath. We will not discuss this
thermalization process here in details. The MD and the system are
reset to their own initial states $\rho_{0}$, which allows a new
cycle to start.

\begin{figure}
\includegraphics[width=8cm]{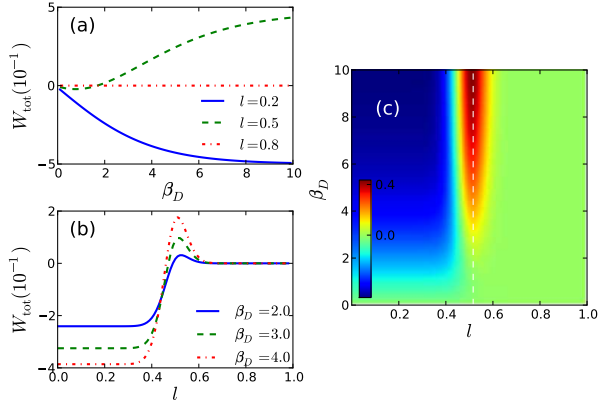} \caption{(Color Online) Work vs insertion position $l$ and MD's inverse temperature
$\beta_{D}$. (a) Total work as a function of $\beta_{D}$ for different
$l=0.2$, $0.5$ and $0.8$. (b)Total work as a function of insertion
position $l$ for different $\beta_{D}=2.0$, $3.0$ and $4.0$. (c)
Contour plot for total work as function of $l$ and $\beta_{D}$.
The position for maximum work extracted is denoted as white dashed
line.}

\label{fig:wtot_gen_imp} 
\end{figure}

\begin{figure}
\includegraphics[width=8cm]{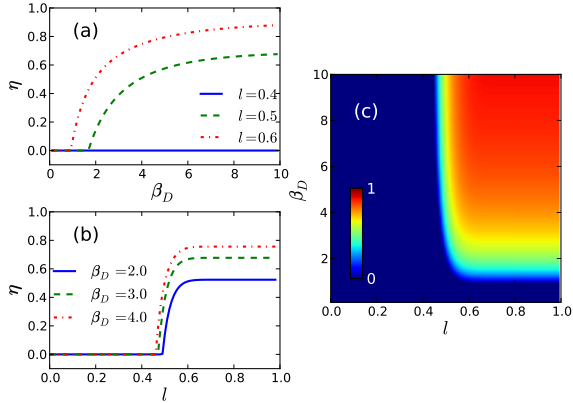} \caption{(Color Online) Efficiency vs insertion position $l$ and O inverse
temperature $\beta_{D}$. (a) Efficiency as a function of $\beta_{D}$
for different $l=0.4$, $0.5$ and $0.6$. (b) Efficiency as a function
$l$ for different $\beta_{D}=2$, $3$ and $4$. (c) Contour plot
of efficiency vs $l$ and $\beta_{D}$.}

\label{fig:eff_gen_imp} 
\end{figure}

\section{Efficiency of Szilard Heat Engine}

For quantum version of the SHE, the quantum coherent based on the
finite size of the chamber results in various different properties
from the classical one. Work is required during the insertion and
removing processes, while the same process can be done freely in the
classical version. The microscopic model here relates the efficiency
of the measurement by MD to the temperature of the heat bath. In the
whole thermodynamic cycle, the work done by the system to outside
is the sum of all the work done in each process, \begin{eqnarray}
W_{\mathrm{tot}} & = & -\left(W_{\mathrm{ins}}+W_{\mathrm{mea}}+W_{\mathrm{exp}}+W_{\mathrm{rev}}\right)\notag\\
 & = & T\left[\left(p_{e}\ln p_{e}+p_{g}\ln p_{g}\right)\right.\notag\\
 &  & -\left(P_{L}p_{g}+P_{R}p_{e}\right)\ln\left(P_{L}p_{g}+P_{R}p_{e}\right)\notag\\
 &  & \left.-\left(P_{L}p_{e}+P_{R}p_{g}\right)\ln\left(P_{L}p_{e}+P_{R}p_{g}\right)\right]\notag\\
 &  & -P_{R}\left(p_{g}-p_{e}\right)\Delta.\end{eqnarray}

To enable the system to do work outside, the temperature of the MD
should be low enough to make sure $W_{\mathrm{tot}}\geq0$, which
is known as the positive-work condition(PWC) \cite{htquan}. To evaluate
the efficiency of QHE, we need to obtain the heat absorbed from the
high temperature heat bath. Different from the classical one, the
exchange of heat with high temperature source persists in each step.
The total heat absorbed from the high temperature source is the sum
over that of all the four steps, \begin{eqnarray}
Q_{\mathrm{tot}} & = & -\left(Q_{\mathrm{ins}}+Q_{\mathrm{mea}}+Q_{\mathrm{exp}}+Q_{\mathrm{rev}}\right)\notag\\
 & = & T\left[\left(p_{e}\ln p_{e}+p_{g}\ln p_{g}\right)\right.\notag\\
 &  & -\left(P_{L}p_{g}+P_{R}p_{e}\right)\ln\left(P_{L}p_{g}+P_{R}p_{e}\right)\notag\\
 &  & \left.-\left(P_{L}p_{e}+P_{R}p_{g}\right)\ln\left(P_{L}p_{e}+P_{R}p_{g}\right)\right].\end{eqnarray}
 Here, the absorbed energy is used to perform work to the outside,
while only the measurement process wastes $W_{\mathrm{mea}}$, which
is released to the low temperature heat bath. It is very interesting
to notice that $W_{\mathrm{mea}}\rightarrow0$ as $\Delta\rightarrow0$,
while the total heat $Q_{\mathrm{tot}}\rightarrow0$ and $W_{\mathrm{tot}}\rightarrow0$.
To check the validity of SLoT, one should concern the the efficiency
of this heat engine in a cycle,

\begin{eqnarray}
\eta & = & 1-\frac{P_{R}\left(p_{g}-p_{e}\right)\Delta}{Q_{\mathrm{tot}}}.\end{eqnarray}

As an example, we consider the special case $l=L/2$, which is similar
to the case of the ordinary SHE with the piston inserted in the center
of the chamber. In this special case, the probabilities for the single
molecule staying at the two sides are the same as that of the classical
one, namely $P_{L}=P_{R}=1/2$. The total work extracted here can
be written in a simple form \begin{equation}
W_{\mathrm{tot}}=T\left(\ln2+p_{e}\ln p_{e}+p_{g}\ln p_{g}\right)-\left(p_{g}-p_{e}\right)\Delta/2.\end{equation}
 In this special case, to make the system capable to do work on the
outside, there is a requirement to the temperature of the demon (low
temperature bath). For example, when we choose $\beta=1$ and $\Delta=0.5$,
the PWC is $\beta_{D}\geq2.09$. This requirement is more strict than
that of Carnot heat engine, $\beta_{D}>1$. And the efficiency of
this heat engine reads \begin{equation}
\eta=1-\frac{\left(p_{g}-p_{e}\right)\Delta}{2T\left(\ln2+p_{e}\ln p_{e}+p_{g}\ln p_{g}\right)},\end{equation}
 which is lower than the corresponding Carnot efficiency \[
\eta_{\mathrm{Carnot}}=1-\frac{T_{D}}{T}.\]
 Here, the efficiency is a monotonic function of the energy spacing
$\Delta$ and reaches its maximum \[
\eta_{\mathrm{max}}=1-\frac{2T_{D}}{T}\leq\eta_{\mathrm{Carnot}}\]
 with $\Delta=0$.

In the general case, we show the work done by the system and efficiency
of the heat engine vs the position of the wall $l$ and the temperature
of demon $\beta_{D}$ in Fig. \ref{fig:wtot_gen_imp} and Fig. \ref{fig:eff_gen_imp}.
As illustrated in Fig. \ref{fig:wtot_gen_imp}(a), for small insertion
position, e.g. $l=0.16$ and $0.36$, the system can not extract positive
work. There exists a critical insertion position $l_{\mathrm{cri}}$
to extract positive work, namely \begin{equation}
T\left(P_{L}^{\mathrm{cri}}\ln P_{L}^{\mathrm{cri}}+P_{R}^{\mathrm{cri}}\ln P_{R}^{\mathrm{cri}}\right)+P_{R}^{\mathrm{cri}}\Delta=0,\end{equation}
 where $P_{R}^{\mathrm{cri}}=P_{R}\left(l_{\mathrm{cri}}\right)$
and $P_{L}^{\mathrm{cri}}=P_{L}\left(l_{\mathrm{cri}}\right)$. This
critical value of insertion position here is $l_{\mathrm{cri}}=0.447$
for the typical parameter chosen here. Due to the requirement of work
in the measurement process, the work extracted is not a symmetric
function of the insertion piston $l$, namely $W_{\mathrm{tot}}\left(0.5-l\right)\neq W_{\mathrm{tot}}\left(0.5+l\right)$,
as illustrated in Fig. \ref{fig:wtot_gen_imp}(b,c). Since the high
energy state $\left|e\right\rangle $ of the demon is utilized to
register the right side for single molecule, more work is need when
$l<L/2$. Due to the requirement of work done by outside agent in
the measurement process, the optimal position to extract maximum work
is not at the center of the potential. The maximum work can be extracted
for a given MD's inverse temperature is reached, when \begin{equation}
\frac{P_{L}^{\mathrm{wmax}}p_{e}+P_{R}^{\mathrm{wmax}}p_{g}}{P_{L}^{\mathrm{wmax}}p_{g}+P_{R}^{\mathrm{wmax}}p_{e}}=e^{-\beta\Delta},\end{equation}
 where $P_{L}^{\mathrm{wmax}}=P_{L}\left(l_{\mathrm{wmax}}\right)$
and $P_{R}^{\mathrm{wmax}}=P_{R}\left(l_{\mathrm{wmax}}\right)$.
It is clear that the position for the maximum work depends on the
temperature of the demon $\beta_{D}$.

In Fig. \ref{fig:eff_gen_imp}, we show the efficiency of this single
molecular heat engine. We consider only the positive work situation,
and set efficiency as $0$ for all the negative work area. Fig. \ref{fig:eff_gen_imp}(a)
shows the monotonous behavior of efficiency as the MD's inverse temperature.
Efficiency is also a monotonous function of the insertion position
$l$, illustrated in Fig. \ref{fig:eff_gen_imp}(b,c), which is not
similar to the total work extracted. It worth noticing that the efficiency
reaches its maximum at $l=1$, while no work can be extracted. Since
the measurement is the only way of wasting energy, it is the only
way to improve the efficiency by reducing $W_{\mathrm{mea}}$ with
decreasing $P_{R}$. The efficiency of QHE reaches the well-known
Carnot efficiency $\eta_{\mathrm{Carnot}}$, when $P_{R}=0$. At the
same time, the total work extracted approaches to zero, namely $W_{\mathrm{tot}}=0$.
We meet this dilemma, since the measurement results in an imperfect
correlation between MD and the system.

Before concluding this paper, we draw our attention to two limit processes\cite{jqliao2009}
again \begin{align}
\lim_{\beta_{D}\rightarrow+\infty}\lim_{\Delta\rightarrow0}\rho_{D} & =\left(\left\vert g\right\rangle \left\langle g\right\vert +\left\vert e\right\rangle \left\langle e\right\vert \right)/2,\\
\lim_{\Delta\rightarrow0}\lim_{\beta_{D}\rightarrow+\infty}\rho_{D} & =\left\vert g\right\rangle \left\langle g\right\vert .\end{align}
 Note that taking the two limits in different orders leads to completely
different results, the latter being a reflection of the spontaneous
symmetry breaking phenomenon. This difference for the MD's initial
state results in the different work extracted, namely, \begin{align}
\lim_{\beta_{D}\rightarrow+\infty}\lim_{\Delta\rightarrow0}W_{\mathrm{tot}} & =0,\\
\lim_{\Delta\rightarrow0}\lim_{\beta_{D}\rightarrow+\infty}W_{\mathrm{tot}} & =k_{B}T\ln2.\end{align}
 The former one means that MD actually gets no information about the
position of molecule and extracts no work, while the latter one show
that MD obtains the exact information on the position of the molecule
and enables the system to perform maximum work to the outside agent.
The same phenomenon has also been revealed in the process of dynamic
thermalization\cite{jqliao2009}.

\section{Conclusions}

In summary, we have studied a quantum version of SHE with a quantum
MD with lower finite temperature than that of the system. We overall
simplified the MD as a two-level system, which carries out measurement
in quantum fashion and controlling the system to do work to the out-side
agent. In this sense, the MD assisted thermodynamic cycle are clarified
as the four steps, insertion, measurement, expansion and removing,
which are all described in the framework of quantum mechanics. In
each step, we also consider the special case to restore the well-known
results in classical version of SHE. We explicitly analyzed the total
work extracted and the corresponding efficiency. To resolve the MD
paradox, we compared the obtained efficiency of the heat engine with
that of Carnot heat engine. It is found the efficiency is always below
that of Carnot since the quantum MD is included as the a part of the
the whole work substance and its functions are also correctly {}``quantized''.
Thus nothing violates the SLoT.

In comparison with the classical version of SHE, the following quantum
natures were discovered in the quantum thermodynamic cycles: (1) The
finite size effect of the potential well was found as reason for the
non-vanishing work required in the insertion and removing of the middle
walls, while the corresponding manipulations could be achieved freely
in the classical case; (2) The quantum coherence is allowed to exist
in the MD's density matrix. It is the decrease of effective temperature
caused by this coherence that actually improves the efficiency of
SHE; (3) In the measurement process, the finite temperature of MD
actually results in the incorrect decision to control the single molecule's
motion. This incorrectness decreased the MD's ability to extract work.
To our best knowledge, even for in the classical case, the similar
investigation has never been carried out; (4) In the whole thermodynamic
cycle, the removing process is necessary in returning to the initial
state for the whole work substance. This fact is neglected in the
previous studies even for the classical SHE.

Finally, we should stress that the model studied here could help to
resolve many paradoxical observations due to heuristic arguments with
hybridization of classical-quantum points of views about thermodynamics.
For instance, it could be recognized that the conventional argument
about the MD paradox only concerns a classical version of MD at the
same temperature as that of the system. Our results can enlighten
the comprehensive understandings about some fundamental problems in
thermodynamics, such as the relationship between quantum unitarity
and SLoT\cite{Dong2010} .

\bigskip{}

\begin{acknowledgments}
HD would like to thank J.N Zhang for helpful discussion. This work
was supported by NSFC through grants 10974209 and 10935010 and by
the National 973 program (Grant No. 2006CB921205). 
\end{acknowledgments}

\section*{Appendix}

In this appendix, we present a detailed calculation for the work done
and efficiency of SHE. Following the calculations \ for the four
steps listed \ in the context step by step, the reader can deeply
understand the physical essences of the MD in some subtle fashion.

\textbf{Step 1: Insertion. }In this process, the changes of internal
energy $\Delta U_{\mathrm{int}}=\mathrm{Tr}\left[\left(H+H_{D}\right)\left(\rho_{\mathrm{ins}}-\rho_{0}\right)\right]$
and total entropy $\Delta S_{\mathrm{ins}}=\mathrm{Tr}\left[-\rho_{\mathrm{ins}}\ln\rho_{\mathrm{ins}}\right]-\mathrm{Tr}\left[-\rho_{\mathrm{0}}\ln\left(\rho_{0}\right)\right]$
is explicitly given by \begin{align}
\Delta U_{\mathrm{int}} & =\sum_{n}p_{n}\left(l\right)E_{n}\left(l\right)+\notag\\
 & \sum_{n}p_{n}\left(L^{\prime}\right)E_{n}\left(L^{\prime}\right)-\sum_{n}p_{n}\left(L\right)E_{n}\left(L\right)\\
 & =\frac{\partial}{\partial\beta}\left[\ln Z\left(L\right)-\ln\mathcal{Z}\left(l\right)\right],\notag\end{align}
 where $L^{\prime}=L-l$ and \begin{align}
\Delta S_{\mathrm{ins}} & =\left(\ln\mathcal{Z}\left(l\right)-\ln Z\left(L\right)\right)+\notag\\
 & \beta\sum_{n}\left[\begin{array}{c}
p_{n}\left(l\right)E_{n}\left(l\right)\\
+p_{n}\left(L^{\prime}\right)E_{n}\left(L^{\prime}\right)-p_{n}\left(L\right)E_{n}\left(L\right)\end{array}\right]\notag\\
 & =\left(1-\beta\frac{\partial}{\partial\beta}\right)\left(\ln\mathcal{Z}\left(l\right)-\ln Z\left(L\right)\right),\end{align}
 where\[
p_{n}\left(y\right)=\frac{\exp\left(-\beta E_{n}\left(y\right)\right)}{Z\left(y\right)}.\]
 For the isothermal process, the work done by outside agent and the
heat exchange are simply $W_{\mathrm{ins}}=\Delta U_{\mathrm{int}}-T\Delta S_{\mathrm{ins}}$
and $Q_{\mathrm{ins}}=-T\Delta S_{\mathrm{ins}}$, namely, \begin{align}
W_{\mathrm{ins}} & =T\left[\ln Z\left(L\right)-\ln\mathcal{Z}\left(l\right)\right],\\
Q_{\mathrm{ins}} & =\left(T-\frac{\partial}{\partial\beta}\right)\left[\ln Z\left(L\right)-\ln\mathcal{Z}\left(l\right)\right].\end{align}

\textbf{Step2: Measurement}. The measurement is realized by a controlled-NOT
unitary operation, which has been illustrated clearly in the Sec.
II. After the measurement process, the density matrix for the total
system is

\begin{eqnarray*}
\rho_{\mathrm{mea}} & = & \left[P_{L}p_{g}\rho^{L}\left(l\right)+P_{R}p_{e}\rho^{R}\left(L^{\prime}\right)\right]\otimes\left\vert g\right\rangle \left\langle g\right\vert \\
 &  & +\left[P_{L}p_{e}\rho^{L}\left(l\right)+P_{R}p_{g}\rho^{R}\left(L^{\prime}\right)\right]\otimes\left\vert e\right\rangle \left\langle e\right\vert .\end{eqnarray*}

The entropy is not changed in this step. And the work done by outside
is \begin{equation}
W_{\mathrm{mea}}=\Delta U_{\mathrm{mea}}=P_{R}\left(p_{g}-p_{e}\right)\Delta.\end{equation}

\textbf{Step3: Controlled expansion}. At the ending of expansion,
the state for the total system reads\begin{eqnarray*}
\rho_{\mathrm{exp}} & = & \left[P_{L}p_{g}\rho^{L}\left(l_{g}\right)+P_{R}p_{e}\rho^{R}\left(L_{g}\right)\right]\otimes\left\vert g\right\rangle \left\langle g\right\vert \\
 &  & +\left[P_{L}p_{e}\rho^{L}\left(L_{e}\right)+P_{R}p_{g}\rho^{R}\left(l_{e}\right)\right]\otimes\left\vert e\right\rangle \left\langle e\right\vert .\end{eqnarray*}
 where $L_{g}=L-l_{g}$ and $L_{e}=L-l_{e}$

We move the wall isothermally. And the work done by out-side agent
can be obtain by the same methods used in insertion process as \begin{widetext}
\begin{eqnarray}
W_{\mathrm{exp}} & = & \mathrm{Tr}\left[\rho_{\mathrm{exp}}\left(H+H_{D}\right)\right]-\mathrm{Tr}\left[\rho_{\mathrm{mea}}\left(H+H_{D}\right)\right]\notag\\
 &  & -T\mathrm{Tr}\left[-\rho_{\mathrm{exp}}\ln\rho_{\mathrm{exp}}\right]+T\mathrm{Tr}\left[-\rho_{\mathrm{mea}}\ln\rho_{\mathrm{mea}}\right]\notag\\
 & = & \sum_{n}[P_{L}p_{g}p_{n}\left(l_{g}\right)E_{n}\left(l_{g}\right)+P_{R}p_{e}p_{n}\left(L_{g}\right)E_{n}\left(L_{g}\right)+P_{L}p_{e}p_{n}\left(L_{e}\right)E_{n}\left(L_{e}\right)\notag\\
 &  & +P_{R}p_{g}p_{n}\left(l_{e}\right)E_{n}\left(l_{e}\right)]+\left(P_{L}p_{e}+P_{R}p_{g}\right)\Delta\notag\\
 &  & -\sum_{n}\left(P_{L}p_{n}\left(l\right)E_{n}\left(l\right)+P_{R}p_{n}\left(L^{\prime}\right)E_{n}\left(L^{\prime}\right)\right)+\left(P_{L}p_{e}+P_{R}p_{g}\right)\Delta\notag\\
 &  & -T\sum_{n}\left[P_{L}p_{g}p_{n}\left(l_{g}\right)\ln P_{L}p_{g}p_{n}\left(l_{g}\right)+P_{R}p_{e}p_{n}\left(L^{\prime}\right)\ln P_{R}p_{e}p_{n}\left(L_{g}\right)\right.\notag\\
 &  & \left.\qquad\qquad+P_{L}p_{e}p_{n}\left(L_{e}\right)\ln P_{L}p_{e}p_{n}\left(L_{e}\right)+P_{R}p_{g}p_{n}\left(l_{e}\right)\ln P_{R}p_{g}p_{n}\left(l_{e}\right)\right]\notag\\
 &  & -T\sum_{n}[p_{g}p_{n}\left(l\right)\ln p_{g}p_{n}\left(l\right)+p_{e}p_{n}\left(L^{\prime}\right)\ln p_{e}p_{n}\left(L^{\prime}\right)\notag\\
 &  & \qquad\qquad+p_{e}p_{n}\left(l\right)\ln p_{e}p_{n}\left(l\right)+p_{g}p_{n}\left(L^{\prime}\right)\ln p_{g}p_{n}\left(L^{\prime}\right)]\\
 & = & P_{L}T\left[\ln Z\left(l\right)-p_{g}\ln Z\left(l_{g}\right)-p_{e}\ln Z\left(L_{e}\right)\right]+P_{R}T\left[\ln Z\left(L^{\prime}\right)-p_{e}\ln Z\left(L_{g}\right)-p_{g}\ln Z\left(l_{e}\right)\right].\end{eqnarray}
 The internal energy changes can be also evaluated as

\begin{eqnarray}
\Delta U_{\mathrm{exp}} & = & \sum_{n}\left[P_{L}p_{g}p_{n}\left(l_{g}\right)E_{n}\left(l_{g}\right)+P_{R}p_{e}p_{n}\left(L_{g}\right)E_{n}\left(L_{g}\right)+P_{L}p_{e}p_{n}\left(L_{e}\right)E_{n}\left(L_{e}\right)+P_{R}p_{g}p_{n}\left(l_{e}\right)E_{n}\left(l_{e}\right)\right]\notag\\
 &  & +\left(P_{L}p_{e}+P_{R}p_{g}\right)\Delta\notag\\
 &  & -\sum_{n}\left(P_{L}p_{n}\left(l\right)E_{n}\left(l\right)+P_{R}p_{n}\left(L'\right)E_{n}\left(L'\right)\right)+\left(P_{L}p_{e}+P_{R}p_{g}\right)\Delta\notag\\
 & = & \sum_{n}\left[P_{L}p_{g}p_{n}\left(l_{g}\right)E_{n}\left(l_{g}\right)+P_{R}p_{e}p_{n}\left(L_{g}\right)E_{n}\left(L_{g}\right)+P_{L}p_{e}p_{n}\left(L_{e}\right)E_{n}\left(L_{e}\right)+P_{R}p_{g}p_{n}\left(l_{e}\right)E_{n}\left(l_{e}\right)\right]\notag\\
 &  & -\sum_{n}\left[P_{L}p_{n}\left(l\right)E_{n}\left(l\right)+P_{R}p_{n}\left(L^{\prime}\right)E_{n}\left(L^{\prime}\right)\right]\notag\\
 & = & P_{L}\frac{\partial}{\partial\beta}\left[\ln Z\left(l\right)-p_{g}\ln Z\left(l_{g}\right)-p_{e}\ln Z\left(L_{e}\right)\right]+P_{R}\frac{\partial}{\partial\beta}\left[\ln Z\left(L^{\prime}\right)-p_{e}\ln Z\left(L_{g}\right)-p_{g}\ln Z\left(l_{e}\right)\right].\end{eqnarray}

\end{widetext}

Then, we obtain the heat exchanges in this process as $Q_{\mathrm{exp}}=-T\Delta S_{\mathrm{exp}}=W_{\mathrm{exp}}-\Delta U_{\mathrm{exp}}$
or

\begin{eqnarray*}
Q_{\mathrm{exp}} & = & P_{L}\left(T-\frac{\partial}{\partial\beta}\right)\left[\ln Z\left(l\right)-p_{g}\ln Z\left(l_{g}\right)-p_{e}\ln Z\left(L_{g}\right)\right]\\
 &  & +P_{R}\left(T-\frac{\partial}{\partial\beta}\right)\left[\ln Z\left(L^{\prime}\right)-p_{e}\ln Z\left(L_{g}\right)-p_{g}\ln Z\left(l_{e}\right)\right].\end{eqnarray*}

\textbf{Step4: Removing}. The piston is removed in this process. After
that, the system returns to its initial state and is not entangled
with MD as The last step would be remove the wall in the trap. The
system is on the state as \begin{eqnarray}
\rho_{\mathrm{rev}} & = & \sum_{n}\frac{\exp\left[-\beta E_{n}\left(L\right)\right]}{Z(L)}\left\vert \psi_{n}\left(L\right)\right\rangle \left\langle \psi_{n}\left(L\right)\right\vert \otimes\\
 &  & \left[\left(P_{L}p_{g}+P_{R}p_{e}\right)\left\vert g\right\rangle \left\langle g\right\vert +\left(P_{L}p_{e}+P_{R}p_{g}\right)\left\vert e\right\rangle \left\langle e\right\vert \right].\notag\end{eqnarray}
 Then, the work done and the heat absorbed \ is respectively \begin{eqnarray}
W_{\mathrm{rev}} & = & \mathrm{Tr}\left[\rho_{\mathrm{rev}}\left(H+H_{D}\right)\right]-\mathrm{Tr}\left[\rho_{\mathrm{exp}}\left(H+H_{D}\right)\right]\\
 &  & -T\mathrm{Tr}\left[-\rho_{\mathrm{rev}}\ln\rho_{\mathrm{rev}}\right]+T\mathrm{Tr}\left[-\rho_{\mathrm{exp}}\ln\rho_{\mathrm{exp}}\right]\notag\end{eqnarray}

\begin{widetext} or \begin{eqnarray}
W_{\mathrm{rev}} & = & \sum_{n}p_{n}\left(L\right)E_{n}\left(L\right)+\left(P_{L}p_{e}+P_{R}p_{g}\right)\Delta\notag\\
 &  & -\sum_{n}\left[P_{L}p_{g}p_{n}\left(l_{g}\right)E_{n}\left(l_{g}\right)+P_{R}p_{e}p_{n}\left(L_{g}\right)E_{n}\left(L_{g}\right)+P_{L}p_{e}p_{n}\left(L_{e}\right)E_{n}\left(L_{e}\right)+P_{R}p_{g}p_{n}\left(l_{e}\right)E_{n}\left(l_{e}\right)\right]\notag\\
 &  & -\left(P_{L}p_{e}+P_{R}p_{g}\right)\Delta\notag\\
 &  & +T[\sum_{n}p_{n}\left(L\right)\ln p_{n}\left(L\right)+\left(P_{L}p_{g}+P_{R}p_{e}\right)\ln\left(P_{L}p_{g}+P_{R}p_{e}\right)+\left(P_{L}p_{e}+P_{R}p_{g}\right)\ln\left(P_{L}p_{e}+P_{R}p_{g}\right)]\notag\\
 &  & -T\sum_{n}\left\{ P_{L}p_{g}p_{n}\left(l_{g}\right)\ln\left[P_{L}p_{g}p_{n}\left(l_{g}\right)\right]+P_{R}p_{e}p_{n}\left(L_{g}\right)\ln\left[P_{R}p_{e}p_{n}\left(L_{g}\right)\right]\right.\notag\\
 &  & \left.\qquad\qquad+P_{L}p_{e}p_{n}\left(L_{e}\right)\ln\left[P_{L}p_{e}p_{n}\left(L_{e}\right)\right]+P_{R}p_{g}p_{n}\left(l_{e}\right)\ln\left[P_{R}p_{g}p_{n}\left(l_{e}\right)\right]\right\} \notag\\
 & = & T\left[-\ln Z\left(L\right)+\left(P_{L}p_{g}+P_{R}p_{e}\right)\ln\left(P_{L}p_{g}+P_{R}p_{e}\right)+\left(P_{L}p_{e}+P_{R}p_{g}\right)\ln\left(P_{L}p_{e}+P_{R}p_{g}\right)\right.\notag\\
 &  & \qquad-P_{L}\ln P_{L}-P_{R}\ln P_{R}-p_{e}\ln p_{e}-p_{g}\ln p_{g}\notag\\
 &  & \left.\qquad+P_{L}p_{g}\ln Z\left(l_{g}\right)+P_{R}p_{e}\ln Z\left(L_{g}\right)+P_{L}p_{e}\ln Z\left(L_{e}\right)+P_{R}p_{g}\ln Z\left(l_{e}\right)\right].\end{eqnarray}
 and \begin{eqnarray}
Q_{\mathrm{rev}} & = & -T\mathrm{Tr}\left[-\rho_{\mathrm{rev}}\ln\rho_{\mathrm{rev}}\right]+T\mathrm{Tr}\left[-\rho_{\mathrm{exp}}\ln\rho_{\mathrm{exp}}\right]\notag\\
 & = & T\left[\sum_{n}p_{n}\left(L\right)\ln p_{n}\left(L\right)+\left(P_{L}p_{g}+P_{R}p_{e}\right)\ln\left(P_{L}p_{g}+P_{R}p_{e}\right)+\left(P_{L}p_{e}+P_{R}p_{g}\right)\ln\left(P_{L}p_{e}+P_{R}p_{g}\right)\right]\notag\\
 &  & -T\sum_{n}\left\{ P_{L}p_{g}p_{n}\left(l_{g}\right)\ln\left[P_{L}p_{g}p_{n}\left(l_{g}\right)\right]+P_{R}p_{e}p_{n}\left(L_{g}\right)\ln\left[P_{R}p_{e}p_{n}\left(L_{g}\right)\right]\right.\notag\\
 &  & \left.\qquad\qquad+P_{L}p_{e}p_{n}\left(L_{e}\right)\ln\left[P_{L}p_{e}p_{n}\left(L_{e}\right)\right]+P_{R}p_{g}p_{n}\left(l_{e}\right)\ln\left[P_{R}p_{g}p_{n}\left(l_{e}\right)\right]\right\} \notag\\
 & = & T\left\{ -\ln Z\left(L\right)+\left(P_{L}p_{g}+P_{R}p_{e}\right)\ln\left(P_{L}p_{g}+P_{R}p_{e}\right)+\left(P_{L}p_{e}+P_{R}p_{g}\right)\ln\left(P_{L}p_{e}+P_{R}p_{g}\right)\right.\notag\\
 &  & \left.\qquad-P_{L}\ln P_{L}-P_{R}\ln P_{R}-p_{e}\ln p_{e}-p_{g}\ln p_{g}+P_{L}p_{g}\ln Z\left(l_{g}\right)+P_{R}p_{e}\ln Z\left(L_{g}\right)+P_{L}p_{e}\ln Z\left(L_{e}\right)+P_{R}p_{g}\ln Z\left(l_{e}\right)\right\} \notag\\
 &  & -\sum_{n}\left[p_{n}\left(L\right)E_{n}\left(L\right)-P_{L}p_{g}p_{n}\left(l_{g}\right)E_{n}\left(l_{g}\right)-P_{R}p_{e}p_{n}\left(L_{g}\right)E_{n}\left(L_{g}\right)\right.\notag\\
 &  & \left.\qquad\qquad-P_{L}p_{e}p_{n}\left(L_{e}\right)E_{n}\left(L_{e}\right)-P_{R}p_{g}p_{n}\left(l_{e}\right)E_{n}\left(l_{e}\right)\right]\notag\\
 & = & T\left[\left(P_{L}p_{g}+P_{R}p_{e}\right)\ln\left(P_{L}p_{g}+P_{R}p_{e}\right)+\left(P_{L}p_{e}+P_{R}p_{g}\right)\ln\left(P_{L}p_{e}+P_{R}p_{g}\right)-P_{L}\ln P_{L}-P_{R}\ln P_{R}-p_{e}\ln p_{e}-p_{g}\ln p_{g}\right]\notag\\
 &  & -\left(T-\frac{\partial}{\partial\beta}\right)\ln Z\left(L\right)+P_{L}p_{g}\left(T-\frac{\partial}{\partial\beta}\right)\ln Z\left(l_{g}\right)+P_{R}p_{e}\left(T-\frac{\partial}{\partial\beta}\right)\ln Z\left(L_{g}\right)\notag\\
 &  & +P_{L}p_{e}\left(T-\frac{\partial}{\partial\beta}\right)\ln Z\left(L_{e}\right)+P_{R}p_{g}\left(T-\frac{\partial}{\partial\beta}\right)\ln Z\left(l_{e}\right).\end{eqnarray}

\end{widetext} The total work extracted by outside agent is the sum
of work extracted in each step as

\begin{align}
W_{\mathrm{tot}} & =-\left(W_{\mathrm{ins}}+W_{\mathrm{mea}}+W_{\mathrm{exp}}+W_{\mathrm{rev}}\right)\notag\\
 & =T[\left(p_{e}\ln p_{e}+p_{g}\ln p_{g}\right)-\left(P_{L}p_{g}+P_{R}p_{e}\right)\ln\left(P_{L}p_{g}+P_{R}p_{e}\right)\notag\\
 & \qquad\qquad-\left(P_{L}p_{e}+P_{R}p_{g}\right)\ln\left(P_{L}p_{e}+P_{R}p_{g}\right)]-P_{R}\left(p_{g}-p_{e}\right)\Delta.\end{align}
 The total heat absorbed can also be obtained as \begin{align}
Q_{\mathrm{tot}} & =-\left(Q_{\mathrm{ins}}+Q_{\mathrm{exp}}+Q_{\mathrm{rev}}\right)\notag\\
 & =T\left[\begin{array}{c}
\left(p_{e}\ln p_{e}+p_{g}\ln p_{g}\right)\\
-\left(P_{L}p_{g}+P_{R}p_{e}\right)\ln\left(P_{L}p_{g}+P_{R}p_{e}\right)\\
-\left(P_{L}p_{e}+P_{R}p_{g}\right)\ln\left(P_{L}p_{e}+P_{R}p_{g}\right)\end{array}\right].\end{align}

\end{document}